\documentclass[10pt,preprint,showpacs,nofootinbib,prd,secnumarabic]{revtex4}

\usepackage{graphicx}
\usepackage{amsmath}
\usepackage{amssymb}
\usepackage{dcolumn}
\usepackage{bm}
\usepackage{hyperref}
\usepackage{comment}

\begin{document}
\renewcommand{\baselinestretch}{1.3}
\newcommand{\be}{\begin{equation}}
\newcommand{\ee}{\end{equation}}
\newcommand{\ba}{\begin{eqnarray}}
\newcommand{\ea}{\end{eqnarray}}
\hoffset=0.0cm
\voffset=-1.0cm
\textheight=20.0cm
\textwidth=15.5cm

\bigskip

\bigskip

\vspace{2cm}
\title{Production of radially excited charmonium mesons in two-body nonleptonic $B_c$ decays}
\vskip 6ex
\author{I. Bediaga}
\email{bediaga@cbpf.br}
\affiliation{Centro Brasileiro de Pesquisas Fisicas, Rua Xavier Sigaud 150, 22290-180, Rio de Janeiro, RJ, Brazil}
\author{J. H. Mu\~noz}
\email{jhmunoz@ut.edu.co}
\affiliation{Departamento de F\'\i sica, Universidad del Tolima, Apartado A\'ereo
546, Ibagu\'e, Colombia\\
Centro Brasileiro de Pesquisas Fisicas, Rua Xavier Sigaud 150, 22290-180, Rio de Janeiro, RJ, Brazil}

\bigskip

\bigskip

\begin{center}
\begin{abstract}
We have computed  branching ratios of  two-body nonleptonic $B_c \to X_{c\overline{c}}M$ decays, where $X_{c\overline{c}}$ is the  radially excited charmonium  $\eta_c(2S)$ or  $\psi(2S)$ meson, and $M$ is a pseudoscalar ($P$) or a vector ($V$) or an  axial-vector ($A(^3P_1)$) meson. We  have assumed  factorization hypothesis and calculated the form factors in the  ISGW2 quark model. Some of these decays  have branching ratios of the order of $10^{-3} - 10^{-4}$.
\end{abstract}
\end{center}


\maketitle
\bigskip

\section{Introduction}

The heavy $B_c$ meson  offers the possibility of studying the two heavy flavors $b$ and $c$ in a meson simultaneously.  It  can only decay through weak interactions and provides a good scenario to study nonleptonic weak decays of heavy mesons. For $B_c$ processes, the contribution of the $c$-quark decays with the $b$-quark being as a spectator  is   $\approx 70\%$ while the $b$-quark decays with the $c$-quark being as a spectator and the weak annihilation decays  contribute  approximately with   $20\%$ and   $10\%$, respectively \cite{Brambilla,Gouz2004}.\\

 The nonleptonic $B_c$ weak decays have been widely studied using different approaches (see the classified bibliography  in Ref. \cite{LXL2010}). The majority of these studies have considered $l=0$ and $l=1$ mesons without radial excitation in final states.  In relation to  excited charmonium states in $B_c$ decays, some works present a systematic  analysis  on  production of orbitally excited charmonium mesons in exclusive nonleptonic and semileptonic $B_c$ decays using different frameworks (see \emph{e.g.}  Refs. \cite{KPS2002,IKS2006,HNV2006,CCWZ2001,WWL2009,ASB2009}). However,  nonleptonic $B_c$ decays with radially excited charmonium mesons  in final state have received less attention in the literature. \\

 At theoretical level,  the observation of a number of new charmoniumlike states above the open charm production threshold \cite{newcharmonium,experiment} has motivated some  works on  production of excited charmonium states in heavy meson decays. For example, recently,  in Ref. \cite{EFG2010} was studied the production of radially and orbitally excited $2P$ and $3S$ charmonium states in semileptonic and nonleptonic $Bc$ decays in the   framework of the relativistic quark model;
 in Ref.  \cite{WL2008} was computed  branching ratios for semileptonic $B_c \to X_{c\overline{c}}l\nu$ decays, where $ X_{c\overline{c}}$ is a radially and orbitally  excited charmonium meson $2S$, $3S$, $4S$, $1P$, $2P$, $1D$, $2D$, $3D$ in  the  light-cone QCD sum rules approach;   and  in Ref. \cite{CFW} was studied the production of excited charmonium states in nonleptonic $B_s$ decays using  generalized factorization together with $SU (3)_F$ symmetry. On the other hand, at experimental level, the high luminosity of the LHC provides the possibility of measuring many decays of the $B_c$ meson \cite{Brambilla,Gouz2004,LHCb}. In particular, some of these $B_c$ channels into charmonium states can be measured at the LHCb experiment where it is expected $\mathcal{O}(10^{9})$ $B_c^+$ mesons with a cross section of 1 $\mu$b and a luminosity of 1 fb$^{-1}$ \cite{He}.  \\

This article is focused on  production of radially excited charmonium $2S$ mesons in two-body nonleptonic weak $B_c$ processes, which arise from  the b-quark decay with the $c$-quark being as a spectator. These  decays are produced by the  $b \to c\overline{q}_iq_j$ transition, where $q_i = u,c$ and $q_j= d,s$. We have omitted the annihilation  contribution because it is expected to be  suppressed, and  assumed naive factorization, which  works reasonably well in two-body nonleptonic $B_c$ decays where the quark-gluon sea is suppressed in the heavy quarkonium \cite{heavyquarkonium}.\\

We have obtained  branching ratios of  two-body nonleptonic  $B_c \to X_{c\overline{c}}(2S)M$ decays, where  $X_{c\overline{c}}(2S)$ is  the  radially excited charmonium  $\eta_c(2S) = \eta_c^{'}$ or $\psi(2S)=\psi^{'}$ meson, and $M$ denotes a pseudoscalar $(P)$ or a vector $(V)$ or an axial-vector $A(^3P_1)$ meson,  using the  ISGW2 quark model \cite{isgw2} for evaluating the $B_c \to \eta_c^{'}$ and the $B_c \to \psi^{'}$ transitions. We have compared our results with previous theoretical predictions obtained in other frameworks based on the relativistic quark model, which works with  the quasipotential approach in quantum field theory \cite{EFG2003}, on the QCD relativistic potential model \cite{CF2000}, on the relativistic constituent quark model based on the Bethe-Salpeter formalism \cite{LT1997}, and on the instantaneous nonrelativistic approximation quark  model \cite{ChCh1994}. For completeness, we have obtained  branching ratios for semileptonic  $B_c \to \eta_c^{'}(\psi^{'})l\nu$ decays and compared with other results obtained in the frameworks mentioned above and in  the light-cone QCD sum rules \cite {WL2008} and QCD sum rules \cite{kiselev2003}.  \\

This paper is organized as follows. In section II, we discuss the weak effective  Hamiltonian and give the form factors for  the $B_c \to \eta_c^{'}$ and $B_c \to \psi^{'}$ transitions.  Numerical results  for branching ratios of nonleptonic and semileptonic $B_c$ decays are presented in section III, and conclusions are given  in section IV.

\bigskip

\section{Hamiltonian and Form Factors}

In this work, we consider only the contribution of current-current operators at tree-level, \textit{i.e}., we do not include penguin diagrams\footnote{It is expected that the  contribution to the decay width of two-body nonleptonic $B_c$ decays from the tree diagram is much larger than the one obtained from the penguin diagrams \cite{penguin}.}. The weak effective  Hamiltonian for the nonleptonic $B_c \to X_{c\overline{c}}(2S)M$ decays, where $X_{c\overline{c}}(2S)$ denotes a radially excited meson $\eta_c^{'}(2\; ^1S_0)$ or $\psi^{'}(2\; ^3S_1)$, and $M$ is a pseudoscalar ($P$) or a vector ($V$) or an axial-vector ($A$) meson, neglecting QCD penguin operators,    is given by
\begin{align}
\mathcal{H}_{eff} &= \frac{G_F}{\sqrt{2}} \{ V_{cb}V_{ud}^*[c_1(\mu)(\overline{c}b)(\overline{d}u) +c_2(\mu)(\overline{d}b)(\overline{c}u)] \notag \\
&  +  V_{cb}V_{cs}^*[c_1(\mu)(\overline{c}b)(\overline{s}c) +c_2(\mu)(\overline{s}b)(\overline{c}c)] \notag  \\
& + V_{cb}V_{us}^*[c_1(\mu)(\overline{c}b)(\overline{s}u) +c_2(\mu)(\overline{s}b)(\overline{c}u)] \notag \\
& + V_{cb}V_{cd}^*[c_1(\mu)(\overline{c}b)(\overline{d}c) +c_2(\mu)(\overline{d}b)(\overline{c}c)]
 \} + h. c.,
\end{align}

where $G_F$ is the Fermi constant, $V_{ij}$ are  CKM factors, ($\overline{q}_{\alpha}q_{\beta}$) is a short notation for the $V - A$ current $\overline{q}_{\alpha}\gamma^{\mu}(1 - \gamma_5)q_{\beta}$, and  $c_{1,2}$ are  the Wilson coefficients. \\

The amplitude of the $B_c \to X_{c\overline{c}}(2S)M$ decay is given by
\begin{equation}
\mathcal{A}(B_c \to X_{c\overline{c}}(2S)M)  =  \left <  X_{c\overline{c}}(2S)M |\mathcal{H}_{eff} | B_c \right > = \frac{G_F}{\sqrt{2}} \sum_i \lambda_i c_i(\mu) \left <  \mathcal{O} \right >_i,
\end{equation}
where $\lambda_i$ is the CKM factor and $\left <  \mathcal{O} \right >_i$ is the matrix element of the local four-quark operators. In the framework of naive factorization, it is assumed that  this element can be approximated by the product of two matrix elements of single currents:
\begin{equation}\nonumber
\langle X_{c\overline{c}}(2S) M | \mathcal{O} |B_c  \rangle_i
\approx   \langle M|J^{\mu}|0  \rangle    \langle  X_{c\overline{c}}(2S)|J_{\mu}|B_c \rangle +  (X_{c\overline{c}}(2S) \leftrightarrow M),
\end{equation}
where $J_{\mu}$ is the weak current. In this way, the hadronic matrix element of a four-quark operator can be expressed as the product of a decay constant and form factors \cite{factorization1, factorization2}. \\

This approach presents a difficulty because the Wilson coefficients, which include the short-distance QCD effects between  $\mu = m_W$ and $\mu = m_b$, are $\mu$ scale and renormalization scheme dependent while $\left <  \mathcal{O} \right >_i$ are $\mu$ scale and renormalization scheme independent. Therefore, the physical amplitude depends on the $\mu$ scale. The naive factorization disentangles the long-distance effects from the short-distance sector assuming that the  matrix element $\left <  \mathcal{O} \right >_i$, at  $\mu$ scale, contain nonfactorizable contributions in order to cancel the $\mu$ dependence and the scheme dependence of $c_i({\mu})$, \textit{i.e.}, this approximation neglects  possible QCD interactions   between the meson $M$ and the $B_cX_{c\overline{c}}$ system \cite{factorization1, factorization2}.  In general, it  works  in some two-body nonleptonic decays of heavy mesons in the limit  of a large number of colours. It is expected that the factorization scheme works reasonably well in two-body nonleptonic $B_c$ decays with radially excited  charmonium mesons in the final state where the quark-gluon sea is supressed in the heavy quarkonium \cite{heavyquarkonium}\footnote{Corrections to  factorization in   the exclusive  $B \to J/\psi + h$ channel, where $h$ is a light meson,  have been studied in Refs. \cite{correctionsCharmomium}. The $B_c \to X_{c\overline{c}}(2S)M$ decays could be  an additional scenario in order to study this type of corrections.}.\\

The Wilson coefficients are related with the QCD coefficients  by means of the expression
\begin{center} $a_{1,2}(\mu) = c_{1,2}(\mu) + \frac{1}{N_c}c_{2,1}(\mu)$. \end{center} In this work, we have assumed large $N_c$ limit to fix the QCD coefficients $a_1 \approx c_1$ and $a_2 \approx c_2$ at $\mu \approx m_b^2$ (there are some works that have assumed this limit such as  \cite{CF2000}, \cite{ChCh1994} and \cite{verma}).\\

We have calculated in the ISGW2 model \cite{isgw2} the form factors of the hadronic matrix elements  $\langle \eta_c^{'}|J_{\mu}|B_c \rangle$ and $\langle \psi^{'}|J_{\mu}|B_c\rangle$. As is well known, this quark model, which  is an improved version of the nonrelativistic ISGW model \cite{isgw}, includes constraints imposed by heavy quark symmetry, relativistic corrections to the matrix elements of the axial vector current and the effective interquark potential, and more realistic polynomial form factors. It is expected that a nonrelativistic treatment of the $B_c$ meson decays with radially excited charmonium mesons provides reliable information \cite{HNV2006, ChCh1994, quarkoniaNR} because both  are heavy quarkonia and these decays arise from  the $b \to c$ transition.  \\

The parametrization of   the $B_c \to \eta_c^{'}$ and $B_c \to \psi^{'}$ transitions are given by \cite{isgw}
\begin{eqnarray}
\langle \eta_c^{'}|J_{\mu}|B_c \rangle &=& f^{'}_{+} \; (p_{B_c}+p_{\eta_c^{'}})_\mu \;  + \; f^{'}_{-} \; (p_{B_c}-p_{\eta_c^{'}})_\mu,  \\
\langle \psi^{'}|J_{\mu}|B_c \rangle &=& ig^{'}\varepsilon_{\mu \nu\rho\sigma}\epsilon^{*\nu}(p_{B_c} + p_{\psi^{'}})^{\rho}(p_{B_c} - p_{\psi^{'}})^{\sigma}  -f^{'}\epsilon^{*}_{\mu}  \nonumber\\
 && - \; (\epsilon^{*}.p_{B_c})[a_+^{'} (p_{B_c} + p_{\psi^{'}})_{\mu} + a_-^{'} (p_{B_c} - p_{\psi^{'}})_{\mu}],
\end{eqnarray}
where $p_{B_c}$, $p_{\eta_c^{'}}$ and $p_{\psi^{'}}$ are the 4-momentum of the $B_c$, $\eta_c^{'}$  and $\psi^{'}$ mesons, respectively, $\epsilon^{*}_{\mu}$ is the polarization of the $\psi^{'}$ meson,  $f^{'}_{+}$,  $f^{'}_{-}$, $f^{'}$, $g^{'}$, $a_+^{'}$ and $a_-^{'}$   are  form factors.

\subsection{Form factors for the $B_c \to \eta_c^{'}$  transition}

The form factors $f_{+}^{'}$ and $f_{-}^{'}$ for  the  $B_c \to \eta_c^{'}$ transition  are given in the ISGW2 model \cite{isgw2} by

\begin{equation}
f_{+}^{'} \; + \; f_{-}^{'} = -\frac{1}{6} \sqrt{\frac{3}{2}}  \frac{\beta^2_{B_c}}{\beta^2_{B_c\eta_c^{'}}}\left( 1 + \frac{m_c}{m_b}  \right) \left[  7 - \frac{\beta^2_{B_c}(5 + \tau)}{\beta^2_{B_c\eta_c^{'}}}  \right]   F_3^{(f_+^{'} \; + \; f_-^{'})},
\end{equation}

\begin{align}
f_{+}^{'} \; - \; f_{-}^{'} &= \sqrt{\frac{3}{2}}  \frac{\tilde{m}_{B_c}}{m_c} \left \{     \left(  \frac{\beta^2_{B_c} - \beta^2_{\eta_c^{'}}}{2\beta^2_{B_c\eta_c^{'}}} + \frac{\tau \beta^2_{B_c}}{3\beta^2_{B_c\eta_c^{'}}}  \right)  +  \right. \notag \\
&  \left. \frac{m_c}{6 \tilde{m}_{\eta_c^{'}}} \frac{\beta^2_{B_c}}{\beta^2_{B_c\eta_c^{'}}}\left( 1 + \frac{m_c}{m_b}  \right) \left[  7 - \frac{\beta^2_{B_c}(5 + \tau)}{\beta^2_{B_c\eta_c^{'}}}      \right ]  \right \}   F_3^{(f_+^{'} \; - \; f_-^{'})},
\end{align}

where

\begin{equation}
F_3^{(f_+^{'} \; \pm \; f_-^{'})} = \left(  \dfrac{\overline{m}_{B_c}}{\tilde{m}_{B_c}} \right )^{\mp\frac{1}{2}}  \left(  \dfrac{\overline{m}_{\eta_c^{'}}}{\tilde{m}_{\eta_c^{'}}} \right )^{\pm \frac{1}{2}}  \left(  \dfrac{\tilde{m}_{\eta_c^{'}}}{\tilde{m}_{B_c}} \right )^{\frac{1}{2}} \left(   \frac{\beta_{B_c}
\beta_{\eta_c^{'}}}{\beta^2_{B_c\eta_c^{'}}}     \right)^{\frac{3}{2}} \left[ 1 + \frac{r^2(t_m - t)}{24}      \right]^{-4},
\end{equation}

 \begin{equation}\label{beta}
  \beta^2_{B_c\eta_c^{'}} = \frac{1}{2}(\beta^2_{B_c} \; + \; \beta^2_{\eta_c^{'}}),
\end{equation}

\begin{equation}\label{tao}
\tau = \frac{m^2_c \beta^2_{\eta_c^{'}}(w - 1)}{\beta^2_{B_c} \beta^2_{B_c\eta_c^{'}}},
\end{equation}

with

 \begin{equation}\label{ere}
 r^2 = \frac{3}{4m_{b}m_{c}} + \frac{3m^2_c}{2\overline{m}_{B_c}\overline{m}_{\eta_c^{'}}\beta^2_{B_c\eta_c^{'}}}
+ \frac{16}{27\overline{m}_{B_c}\overline{m}_{\eta_c^{'}}}ln\left[  \frac{\alpha_s(\mu_{QM})}{\alpha_s(m_c)}  \right],
\end{equation}

\begin{equation}\label{doblev}
w = 1 + \frac{t_m - t}{2\overline{m}_{B_c} \overline{m}_{\eta_c^{'}}}.
\end{equation}

The values of the $\beta$ parameter, which is the relativistic correction to the hyperfine-corrected wave function in the ISGW2 model,   are given in \cite{isgw2}. $t  = (p_{B_c} - p_{\eta_c^{'}})^2 \equiv q^2$ is the momentum transfer, $t_m=(m_{B_c} - m_{\eta_c^{'}})^2$ is the maximum momentum transfer,   $\overline{m}_X$ is the hyperfine-averaged physical mass of the $X$ meson, $\tilde{m}_X$ is the sum of the masses of constituent quarks of the $X$ meson , $\mu_{QM} \approx 1$ GeV is a quark model scale. The momentum transfer $q^2$ is constant  for  the two-body nonleptonic $B_c \to \eta_c^{'}M$ decay: $q^2 = m_M^2$. In Table I, we show the values of $f_{+}^{'}$ and $f_{-}^{'}$ at momentum transfer $q^2 =0, \; t_m$ in the ISGW2 model. Also, in Fig. 1 we plot these form factors in the kinematical range $0 \leq q^2  \leq (m_{B_c} - m_{\eta_c^{'}})^2$.

\begin{figure}[htp]
\centering
$$\begin{array}{cc}
$\includegraphics[width=7cm]{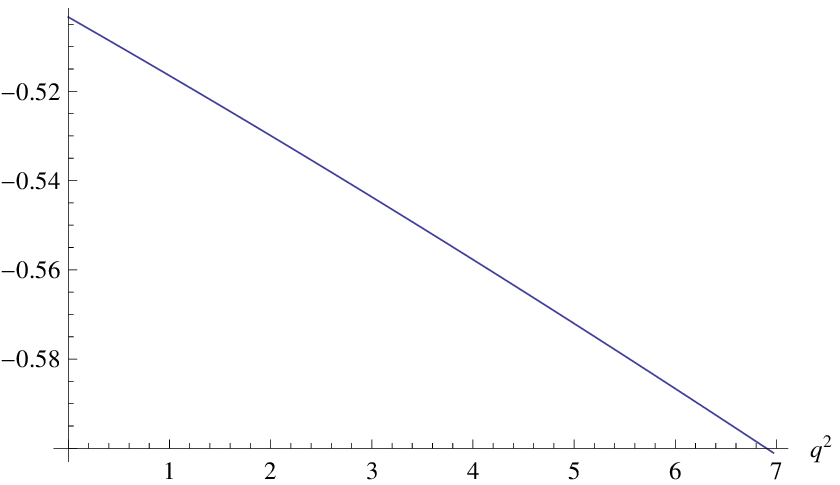}$ & $\includegraphics[width=7cm]{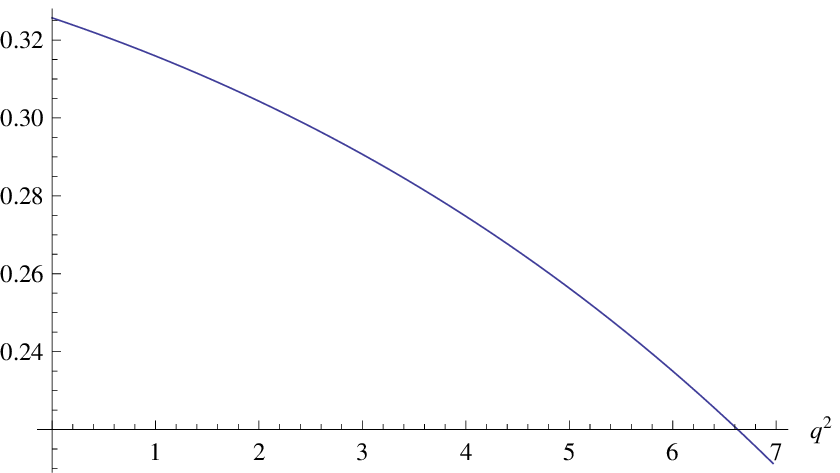}$\\
$(a)$ & $(b)$
\end{array}$$
  \caption{Form factors for the $B_c \to \eta_c'$ transition: (a) $f_{-}'(q^2)$, (b) $f_{+}'(q^2)$.}\label{fig:formapolar}
\end{figure}

\subsection{Form factors for the $B_c \to \psi^{'}$  transition}

The form factors $f^{'}$, $g^{'}$ and $a^{'}_{\pm}$  are given in the ISGW2 model \cite{isgw2} by:

\begin{equation}
f^{'} = (0.899)\sqrt{\dfrac{3}{2}}\tilde{m}_{B_c} (1+w) \left [ \frac{\beta^2_{B_c} - \beta^2_{\psi^{'}}}{2\beta^2_{B_c\psi^{'}}} + \frac{\tau \beta^2_{B_c}}{3\beta^2_{B_c\psi^{'}}} \right ] F_3^{(f^{'})},
\end{equation}

\begin{equation}
g^{'} = \frac{1}{2}\sqrt{\frac{3}{2}}\left[ \left( \frac{1}{m_c}-\frac{m_c\beta^2_{B_c}}{2\mu_-\tilde{m}_{\psi^{'}}\beta^2_{B_c\psi^{'}}}  \right)
\left( \frac{\beta^2_{B_c} - \beta^2_{\psi^{'}}}{2\beta^2_{B_c\psi^{'}}} + \frac{\tau \beta^2_{B_c}}
{3\beta^2_{B_c\psi^{'}}} \right)
+ \frac{m_c\beta^2_{B_c}\beta^2_{\psi^{'}}}{3\mu_-\tilde{m}_{\psi^{'}}\beta^4_{B_c\psi^{'}}}
   \right]  F_3^{(g^{'})},
\end{equation}

\begin{eqnarray} \nonumber
a^{'}_+ \; + \; a^{'}_- &=& -\sqrt{\frac{2}{3}}\frac{\beta^2_{B_c}}{m_cm_b\beta^2_{B_c\psi^{'}}} \left \{
\frac{7m^2_c\beta^4_{\psi^{'}}(1+\frac{\tau}{7})}{8\tilde{m}_{B_c}\beta^4_{B_c\psi^{'}}}
- \frac{5m_c\beta^2_{\psi^{'}}(1 + \frac{\tau}{5})}{4\beta^2_{B_c\psi^{'}}} \right.\\
 && \left.
- \frac{3m^2_c\beta^4_{\psi^{'}}}       {8\tilde{m}_{B_c}\beta^2_{B_c}\beta^2_{B_c\psi^{'}}}
+ \frac{3m_c\beta^2_{\psi^{'}}}{4\beta^2_{B_c}} \right \}
 F_3^{(a^{'}_+ \; + \; a^{'}_-)},
\end{eqnarray}

\begin{eqnarray} \nonumber
a^{'}_+ \; - \; a^{'}_- &=& \sqrt{\frac{2}{3}}  \frac{3\tilde{m}_{B_c}}{2m_b\tilde{m}_{\psi^{'}}}  \left \{
1 - \frac{\beta^2_{B_c}(1+\frac{\tau}{7})}{\beta^2_{B_c\psi^{'}}}
- \frac{m_c\beta^2_{\psi^{'}}}{2\tilde{m}_{B_c}\beta^2_{B_c\psi^{'}}} \left(   1 - \frac{5\beta^2_{B_c}(1 + \frac{\tau}{5})}{3\beta^2_{B_c\psi^{'}}} \right) \right. \\
&& \left. -\frac{7m^2_c\beta^2_{B_c}\beta^2_{\psi^{'}}}{12m_c\tilde{m}_{B_c}\beta^4_{B_c\psi^{'}}} \left( 1 - \frac{\beta^2_{\psi^{'}}}{\beta^2_{B_c\psi^{'}}} + \frac{\tau \beta^2_{B_c}}{7\beta^2_{B_c\psi^{'}}}  \right)
\right \}  F_3^{(a^{'}_+ \; - \; a^{'}_-)},
\end{eqnarray}

where

\begin{equation}
 F_3^{(f^{'})} = \left( \dfrac{\overline{m}_{B_c}}{\tilde{m}_{B_c}}  \right )^{\frac{1}{2}}  \left( \dfrac{\overline{m}_{\psi^{'}}}{\tilde{m}_{\psi^{'}}}   \right )^{\frac{1}{2}}      \left( \frac{\tilde{m}_{\psi^{'}}}{\tilde{m}_{B_c}}  \right)^{\frac{1}{2}}   \left( \frac{\beta_{B_c}\beta_{\psi^{'}}}{\beta^2_{B_c\psi^{'}}}  \right)^{\frac{3}{2}} \left[ 1 + \frac{r^2(t_m - t)}{24} \right]^{-4},
\end{equation}

\begin{equation}
F_3^{(g^{'})} = \left( \dfrac{\overline{m}_{B_c}}{\tilde{m}_{B_c}}  \right )^{-\frac{1}{2}}  \left( \dfrac{\overline{m}_{\psi^{'}}}{\tilde{m}_{\psi^{'}}}   \right )^{-\frac{1}{2}}      \left( \frac{\tilde{m}_{\psi^{'}}}{\tilde{m}_{B_c}}  \right)^{\frac{1}{2}}   \left( \frac{\beta_{B_c}\beta_{\psi^{'}}}{\beta^2_{B_c\psi^{'}}}  \right)^{\frac{3}{2}} \left[ 1 + \frac{r^2(t_m - t)}{24} \right]^{-4},
\end{equation}

\begin{equation}
F_3^{(a^{'}_+ \; + \; a^{'}_-)} =  \left( \frac{\overline{m}_{B_c}}{\tilde{m}_{B_c}}  \right )^{-\frac{3}{2}}  \left( \frac{\overline{m}_{\psi^{'}}}{\tilde{m}_{\psi^{'}}}   \right )^{\frac{1}{2}}      \left( \frac{\tilde{m}_{\psi^{'}}}{\tilde{m}_{B_c}}  \right)^{\frac{1}{2}}   \left( \frac{\beta_{B_c}\beta_{\psi^{'}}}{\beta^2_{B_c\psi^{'}}}  \right)^{\frac{3}{2}} \left[ 1 + \frac{r^2(t_m - t)}{24} \right]^{-4},
\end{equation}

\begin{equation}
F_3^{(a^{'}_+ \; - \; a^{'}_-)} =  \left( \frac{\overline{m}_{B_c}}{\tilde{m}_{B_c}}  \right )^{-\frac{1}{2}}  \left( \frac{\overline{m}_{\psi^{'}}}{\tilde{m}_{\psi^{'}}}   \right )^{-\frac{1}{2}}      \left( \frac{\tilde{m}_{\psi^{'}}}{\tilde{m}_{B_c}}  \right)^{\frac{1}{2}}   \left( \frac{\beta_{B_c}\beta_{\psi^{'}}}{\beta^2_{B_c\psi^{'}}}  \right)^{\frac{3}{2}} \left[ 1 + \frac{r^2(t_m - t)}{24} \right]^{-4},
\end{equation}

\begin{equation}
\mu_{\pm} = \left(  \frac{1}{m_c} \pm \frac{1}{m_b}   \right)^{-1}.
\end{equation}

$\beta^2$, $\tau$, $r^2$ and $w$ are given by Eqs. (\ref{beta}), (\ref{tao}), (\ref{ere}) and (\ref{doblev}), respectively, substituting $\eta_c^{'}$ by $\psi^{'}$. The factor $0.899$ in $f^{'}$ is a relativistic correction to the matrix elements of the axial vector current in the ISGW2 model \cite{isgw2}.\\

In Table I, we show the values of $f^{'}$, $g^{'}$ and $a^{'}_{\pm}$ at momentum transfer $q^2 =0, \; t_m$, evaluated in the ISGW2 model. Moreover, in Fig. 2  we  display these form factors in the kinematical region $0 \leq q^2  \leq (m_{B_c} - m_{\psi^{'}})^2$.

\begin{figure}[htp]
\centering
$$\begin{array}{cc}
$\includegraphics[width=6cm]{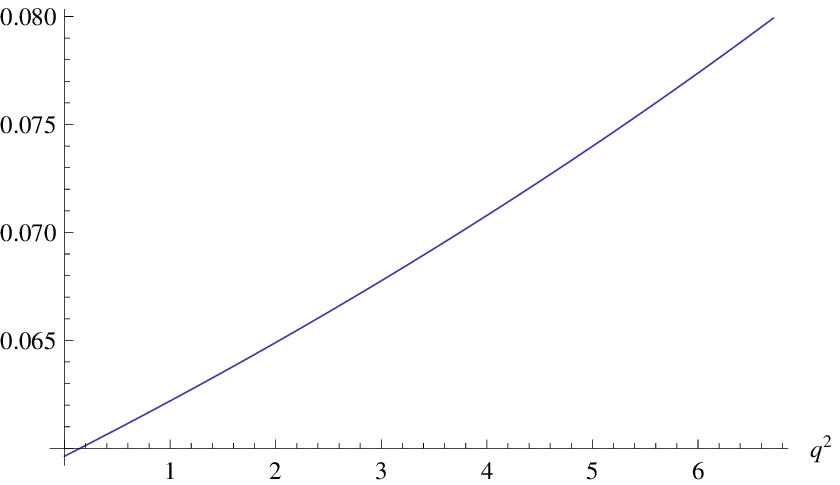}$ & $\includegraphics[width=6cm]{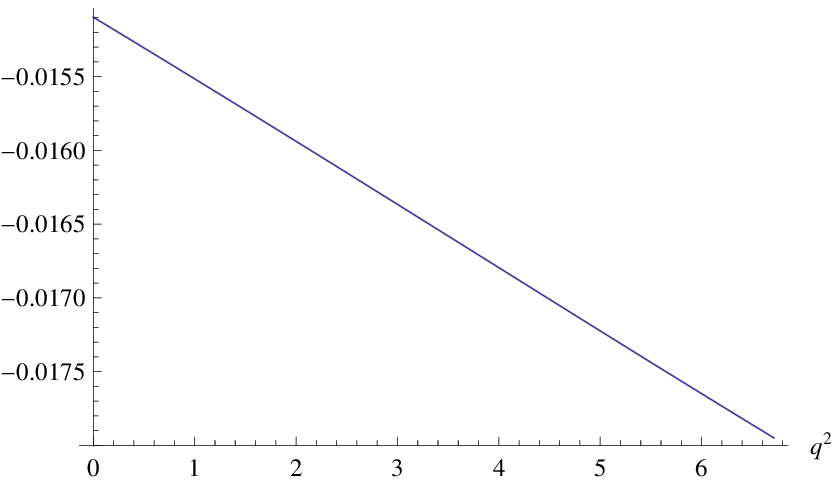}$\\
$(a)$ & $(b)$ \\
$\includegraphics[width=6cm]{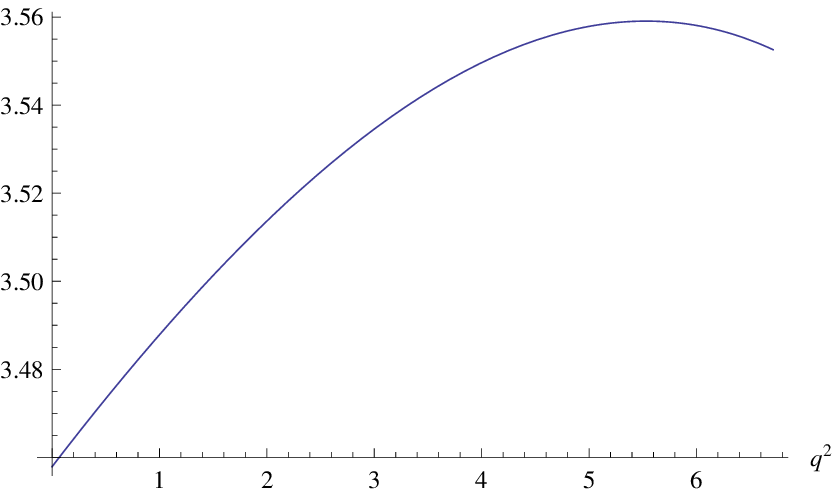}$ & $\includegraphics[width=6cm]{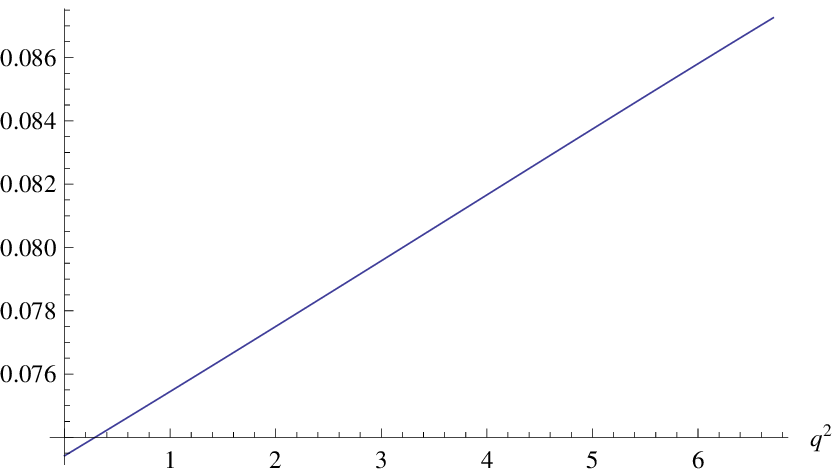}$\\
$(c)$ & $(d)$
\end{array}$$
  \caption{Form factors for the $B_c \to \psi'$ transition: (a) $a_{-}'(q^2)$, (b) $a_{+}'(q^2)$, (c) $f'(q^2)$, (d) $g'(q^2)$.}\label{fig:formfapsi}
\end{figure}

\begin{table}[htp]
{\small Table I. Form factors for the $B_c \to \eta_c^{'}$ and   $B_c \to \psi'$ transitions at $q^2 = 0, \; t_m$ in the ISGW2 model.}
\par
\begin{center}
\renewcommand{\arraystretch}{2}
\begin{tabular}{c|c|c|c|c|c|c}
  \hline\hline
  & $f_{+}^{'}(q^2)$ & $f_{-}^{'}(q^2)$ & $f^{'}(q^2)$ & $g^{'}(q^2)$ & $a_{+}^{'}(q^2)$ & $a_{-}^{'}(q^2)$ \\
  \hline
  $q^2 =0$ & 0.325 & -0.503  &     3.457 & 0.073 &-0.015 & 0.059\\
 $q^2 = t_m$ & 0.211 & -0.601 &     3.552 & 0.087 &-0.017 &0.079 \\
 \hline\hline
\end{tabular}
\end{center}
\end{table}


\bigskip

\section{Numerical values and discussion}

In order to obtain  branching ratios of nonleptonic and semileptonic $B_c$ decays with radially excited charmonium mesons in the final state, we take the meson masses from the PDG \cite{pdg} and the following numerical values:

\begin{itemize}
\item For CKM factors \cite{pdg}: $|V_{cb}|=40.6 \times 10^{-3}$, $|V_{ud}|=0.97425$,  $|V_{cs}|=1.023$, $|V_{us}|=0.2252$, $|V_{cd}|=0.230$.
\item For quark masses (in GeV) \cite{isgw2}: $m_b = 5.2$, $m_c = 1.82$, $m_s=0.55$, $m_u = m_d=0.33$.
\item For QCD coefficients: $a_1 = 1.14$, $a_2=-0.2$ (see for example  Refs. \cite{IKS2006,HNV2006,CF2000,ChCh1994,kiselev2003,verma}).
\item For decay constants (in GeV): $f_{\pi^-}=0.131$ \cite{CCH2004}, $f_{K^-}= 0.160$ \cite{CCH2004}, $f_{D^-}=0.227$ \cite{IKS2006}, $f_{D_s^-}=0.259$ \cite{CLEO}, $f_{\rho^-}=0.216$ \cite{CC2010}, $f_{K^{*-}}=0.210$ \cite{CCH2004}, $f_{D^{*-}}=0.249$ \cite{IKS2006}, $f_{D_s^{*-}}=0.266$ \cite{IKS2006},  $f_{a_1^-}=0.238$ \cite{CC2010}, $f_{K_1(1270)}=-0.170$ \cite{CC2010}, $f_{K_1(1400)}=-0.139$ \cite{CC2010}, $f_{\eta_c'}=0.270$ \cite{CFW}, $f_{\psi^{'}}= 0.304$ \cite{WL2008}.
\item $\beta$ parameters (in GeV) from the ISGW2 model \cite{isgw2}: $\beta_{B_c}=0.92$, $\beta_{\eta_c^{'}}= 0.88$, $\beta_{\psi^{'}}=0.62$, $\beta_{D}=0.45$, $\beta_{D_s}=0.56$, $\beta_{D^*}= 0.38$, $\beta_{D_s^*} = 0.44$.
\item $\tau_{B_c}=0.453 \times 10^{-12}$ s \cite{pdg}.
\end{itemize}

 Expressions for  decay widths of  two-body nonleptonic $B_c \to X_{c\overline{c}}(2S)M$, where $X_{c\overline{c}}(2S) = \eta_c^{'}(2 \; ^1S_0), \; \psi^{'}(2 \; ^3S_1)$ and $M = P, \; V, \; A(^3P_1)$ are well known in the literature (see for example the overview given in Ref. \cite{MQ}).\\

In Table II,  we show our results for the branching ratios of two-body nonleptonic $B_c \to \eta_c^{'}P, \; \eta_c^{'}V,  \; \eta_c^{'}A(^3P_1)$ decays    and compare with  predictions of other approaches based on  relativistic quark models  \cite{EFG2003,CF2000,LT1997}, and on the instantaneous nonrelativistic approximation quark  model \cite{ChCh1994}. We have obtained numerical values of  branching ratios from these references  taking $a_1 =1.12 $  y $a_2=-0.2$.   In general, we can see that branching ratios have close values in all models. Our results agree with  predictions of Ref. \cite{ChCh1994}, except for  $B_c^- \to \eta_c^{'}D^-(D_s^-)$ decays. In this case, our numerical values are smaller than ones obtained in \cite{ChCh1994}. On the other hand,  results obtained in Ref. \cite{CF2000} are smallest for all channels. For the branching ratio of the  $B_c^- \to \eta_c^{'}D^{*-}$ mode there is a remarkable difference between our numerical value and the one obtained in \cite{CF2000}. \\

 We can see, in Table II, that the CKM favored  $B_c ^- \to \eta_c^{'}\pi^-, \; \eta_c^{'}\rho^-, \eta_c^{'}a_1^-, \; \eta_c^{'}D_s^{*-}$ modes have branching ratios of the order of $\approx 10^{-4}$. These branching ratios could be measured in the future at the LHCb experiment. We  also obtain that
\begin{equation}
\frac{Br(B_c \to \eta_c^{'}V(q_i\overline{q}_j))}{Br(B_c \to \eta_c^{'}P(q_i\overline{q}_j))} \gtrsim (1.4 \; -  \; 4.8). \\
\end{equation}
Let us note  that in  Refs.  \cite{CF2000} and \cite{ChCh1994} this quotient gives $< 1$ when $V = D^{*-}, \; D_s^{*-}$ and $P = D^{-}, \; D_s^{-}$, respectively. Therefore, this ratio could offer a test for these quark models.  \\

The $B_c ^- \to \eta_c^{'}D_{(s)}^{(*)-}$ decays  have two contributions: with $W$-external emission (proportional to   QCD coefficient $a_1$) and with $W$-internal emission (proportional to  QCD coefficient $a_2$, which is negative). For the second contribution we  need to evaluate the $B_c \to D(D_s)$ and  $B_c \to D^*(D_s^*)$ transitions.  We  obtained the form factors for these transitions in the ISGW2 model.  It is important to note that the interference term in  $B_c ^- \to \eta_c^{'}D^{*-}$ and $B_c ^- \to \eta_c^{'}D_s^{*-}$ modes is positive because $a_2$ and the form factor $A_0(t = m^2_{\eta_c^{'}})$ (this form factor appears  in the parametrization of $B_c \to V$ transition in Ref. \cite{wsb}) are negative. So, the behavior of the interference term in $B_c \to \eta_c^{'}D(D_s)$ and $B_c \to \eta_c^{'}D^*(D_s^*)$ decays is different: in $B_c \to \eta_c^{'}D(D_s)$ decays the dominant contribution comes from the $W$-external emission term while in  $B_c \to \eta_c^{'}D^*(D_s^*)$ channels, the contributions that arise from the $W$-external emission and the interference term are of the same order.  \\

For the $B_c \to \eta_c^{'}A(^3P_1)$ modes, where $A(^3P_1)$ is an  axial-vector meson we found that the branching ratio of the CKM favored $B_c^- \to \eta_c{'}a_1^-$ decay is of the order of $10^{-4}$ and is smaller that $Br(B_c \to \eta_c^{'}\rho^-)$. In fact,
\begin{equation}\label{mayorUNO}
\frac{Br(B_c \to \eta_c^{'}\rho^-)}{Br(B_c \to \eta_c^{'}a_1^-)} \approx 1.12.
\end{equation}
On the other hand, when we consider the strange  $K_1(1270)$ and $K_1(1400)$ mesons, which are a mixture  of $K_{1A}$ and $K_{1B}$ mesons, it is obtained
\begin{equation}
\frac{Br(B_c \to \eta_c^{'}K_1^-(1270))}{Br(B_c \to \eta_c^{'}K_1^-(1400))} \approx 1.73.
\end{equation}
This quotient  can be an additional  test for the $K_{1A}-K_{1B}$ mixing angle.\\

\begin{table}[ht]
{\small Table II. Branching ratios of the $B_c \to \eta_c^{'}M$ decays, where $M = P, \; V, \; A(^3P_1)$}.
\par
\begin{center}
\renewcommand{\arraystretch}{2}
\begin{tabular}{c|c|c|c|c|c}
  \hline
 Decay & This work & \cite{EFG2003} & \cite{CF2000}& \cite{LT1997}&\cite{ChCh1994} \\
  \hline\hline
  $B_c^- \to \eta_c^{'}\pi^-$ &  $2.4 \times 10^{-4}$ & $1.7 \times 10^{-4} $& $6.6 \times 10^{-5} $&$2.2 \times 10^{-4} $ & $2.4 \times 10^{-4} $ \\
  $B_c^- \to \eta_c^{'}K^-$ & $1.8 \times 10^{-5}$ &$1.25 \times 10^{-5} $ &$4.9 \times 10^{-6} $ &$1.6 \times 10^{-5} $ & $1.8 \times 10^{-5} $ \\
   $B_c^- \to \eta_c^{'}D^-$ & $5.7 \times 10^{-6}$ & & $2.2 \times 10^{-6}$ & & $2 \times 10^{-5}$  \\
   $B_c^- \to \eta_c^{'}D_s^-$ & $6.7 \times 10^{-5}$ & & $7.85 \times 10^{-5}$ & & $8.7 \times 10^{-4}$ \\
  \hline\hline
  $B_c^- \to \eta_c^{'}\rho^-$ & $5.5 \times 10^{-4}$ &$3.6\times 10^{-4} $ &$1.4 \times 10^{-4} $ &$5.25 \times 10^{-4} $ &$5.5 \times 10^{-4} $ \\
  $B_c^- \to \eta_c^{'}K^{*-}$ & $2.6 \times 10^{-5}$ & $1.9 \times 10^{-5} $&$7.15 \times 10^{-6} $ &$2.5 \times 10^{-5} $ &$2.8 \times 10^{-5} $  \\
  $B_c^- \to \eta_c^{'}D^{*-}$ &$2.1 \times 10^{-5} $ & &$7.8 \times 10^{-8} $  & & $1.1 \times 10^{-5} $ \\
$B_c^- \to \eta_c^{'}D_s^{*-}$ &$4.5 \times 10^{-4} $ & &$2 \times 10^{-5} $ & &$4.4 \times 10^{-4} $ \\
\hline\hline
$B_c^- \to \eta_c^{'}a_1^-$ & $4.9 \times 10^{-4}$ & &$1.3 \times 10^{-4} $ & & \\
$B_c^- \to \eta_c^{'}K_1^-(1270)$ & $1.3 \times 10^{-5} $ & & & &\\
$B_c^- \to \eta_c^{'}K_1^-(1400)$ & $7.5 \times 10^{-6}$ & & & &\\
\hline\hline
\end{tabular}
\end{center}
\end{table}

In Table III, we present our predictions for  the branching ratios of  $B_c \to \psi^{'}P, \; \psi^{'}V$,  $\psi^{'}A(^3P_1)$ decays and compare our results with those obtained in other approaches based on relativistic \cite{EFG2003,CF2000,LT1997}, and nonrelativistic quark models \cite{ChCh1994}\footnote{Ref. \cite{Ball2000} summarizes some of these theoretical predictions.}. We have obtained numerical values of the branching ratios from these references  taking $a_1 =1.12 $  y $a_2=-0.2$. Our predictions are the biggest. They are bigger than those obtained in Ref. \cite{CF2000}  and in Ref. \cite{ChCh1994}  approximately by a factor of $(1.93 - 12.11)$ and of $(1.18 - 2.28)$, respectively. \\

We see, in Table III,  that the CKM favored $B_c \to \psi^{'}\rho,  \;  \psi^{'}D_s^*$ and $B_c \to  \psi^{'}a_1$ modes, which are decays of the type $B_c \to V(2S)V(1S)$ and $B_c \to V(2S)A(1S)$, respectively, have branching ratios of the order of $10^{-3}$. The other CKM favored  $B_c \to \psi^{'}\pi, \; \psi^{'}(D_s)$ processes, which are $B_c \to V(2S)P(1S)$ channels,  have branching ratios of the order of $10^{-4}$. In general, we obtain
\begin{equation}
\frac{Br(B_c \to \psi^{'} V(q_i\overline{q}_j))}{Br(B_c \to \psi^{'} P(q_i\overline{q}_j))} \gtrsim 2.
\end{equation}

 For $B_c \to \psi^{'}A(^3P_1)$ decays, where $A(^3P_1)$ denotes an axial-vector meson, we obtain that  the branching ratio of the $B_c^- \to \psi^{'}a_1^-$ channel is the biggest. A similar result it is obtained in Ref. \cite{CF2000}. In fact,
\begin{equation}
\frac{Br(B_c \to \psi^{'}\rho^-)}{Br(B_c \to \psi^{'}a_1^-)} \approx 0.74.
\end{equation}
In this case, this quotient  is $< 1$ while the same ratio changing $\psi^{'}$ by $\eta_c^{'}$ is $> 1$ (see Eq. (\ref{mayorUNO})). On the other hand, when the axial-vector meson is a strange meson, we obtain
\begin{equation}
\frac{Br(B_c \to \psi^{'}K_1^-(1270))}{Br(B_c \to \psi^{'}K_1^-(1400))} \approx 1.5.
\end{equation}
This ratio provides an additional test  for the $K_{1A}-K_{1B}$ mixing angle.\\

 The $B_c \to \psi^{'}D_{(s)}^{(*)}$ decays also have two contributions: one with $W$-external emission and proportional to  QCD coefficient $a_1$ and another with $W$-internal emission and proportional to QCD coefficient $a_2$.  For obtaining the  branching ratios of these processes we need to evaluate the form factors for the $B_c \to D(D_s)$ and the $B_c \to D^*(D_s^*)$ transitions. We computed these form factors in the ISGW2 model. We obtain that in all cases the interference is destructive and  it is smaller in the $B_c \to \psi^{'}D(D_s)$ decays. We remark that Kiselev \cite{kiselev2003} found a similar effect in the interference of two-body nonleptonic $B_c \to X_{c\overline{c}}(1S)D^{(*)}_{(s)}$ decays, where $X_{c\overline{c}}(1S)$ is the $\eta_c$ or the  $J/\psi$ meson, i.e., a charmonium meson without radial excitation. In  Table IX of the first paper of Ref. \cite{kiselev2003} it is showed    the value of the interference term in these decays. \\

From Tables II and  III, we obtain that  $Br(B_c \to \psi^{'}M)  > Br(B_c \to \eta_c^{'}M)$. Specifically, it is found that
\begin{equation}\nonumber
\frac{Br(B_c \to \psi^{'}P)}{Br(B_c \to \eta_c^{'}P)} \approx (1.6 \; - \; 6.3), \ \ \
\frac{Br(B_c \to \psi^{'}V)}{Br(B_c \to \eta_c^{'}V)} \approx (2 \; - \; 3), \ \ \
\frac{Br(B_c \to \psi^{'}A)}{Br(B_c \to \eta_c^{'}A)} \approx (3 \; - \; 3.6).
\end{equation}
This ratio is bigger for those decays that have two contributions. On the other hand, for $B_c \to P(2S)V(1S)$ and $B_c \to V(2S)P(1S)$ decays we obtain
\begin{equation}\nonumber
\frac{Br(B_c \to \eta_c^{'}V(q_i\overline{q}_j))}{Br(B_c \to \psi^{'}P(q_i\overline{q}_j))} \approx 0.8,
\end{equation}
except for $V = \rho^-$ and $P = \pi^-$. In this case, the ratio is $1.44$.\\

\begin{table}[ht]
{\small Table III. Branching ratios of the $B_c \to \psi^{'}M$ decays, where $M = P, \; V, \; A(^3P_1)$}.
\par
\begin{center}
\renewcommand{\arraystretch}{1.5}
\begin{tabular}{c|c|c|c|c|c}
  \hline
    Decay & This work & \cite{EFG2003} & \cite{CF2000}& \cite{LT1997}&\cite{ChCh1994} \\
  \hline\hline
 $B_c^- \to \psi^{'}\pi^-$ & $3.7 \times 10^{-4}$ &$1.1 \times 10^{-4} $ &$2 \times 10^{-4} $ &$6.3 \times 10^{-5} $ &$2.2 \times 10^{-4} $ \\
  $B_c^- \to \psi^{'}K^-$ & $2.9 \times 10^{-5} $ &$8 \times 10^{-6} $ &$8.9 \times 10^{-6} $ & $4.45 \times 10^{-6} $& $1.6 \times 10^{-5} $ \\
  $B_c^- \to \psi^{'}D^-$ & $2.4 \times 10^{-5}$ & &$7.3 \times 10^{-6} $ & &$1.1 \times 10^{-5} $ \\
  $B_c^- \to \psi^{'}D_s^-$ & $5.25 \times 10^{-4}$ & &$1.2 \times 10^{-4} $& &$4.4 \times 10^{-4} $ \\
  \hline\hline
  $B_c^- \to \psi^{'}\rho^-$ & $1.1 \times 10^{-3}$ & $1.8 \times 10^{-4} $& $4.8 \times 10^{-4} $& $1.6 \times 10^{-4} $&$6.3 \times 10^{-4} $ \\
  $B_c^- \to \psi^{'}K^{*-}$ & $5.7 \times 10^{-5}$ &$9.8 \times 10^{-6} $ &$2.7 \times 10^{-5} $ &$8.1 \times 10^{-6} $ &$3.4 \times 10^{-5} $ \\
  $B_c^- \to \psi^{'}D^{*-}$ &$6.3 \times 10^{-5}$ & &$5.2 \times 10^{-6} $ & &\\
$B_c^- \to \psi^{'}D_s^{*-}$ & $1.2 \times 10^{-3}$ & &$1.7 \times 10^{-4} $ & &\\
\hline\hline
$B_c^- \to \psi^{'}a_1^-$ & $1.5 \times 10^{-3}$ & &$5.8 \times 10^{-4} $ & &\\
$B_c^- \to \psi^{'}K_1^-(1270)$ & $4 \times 10^{-5}$ & & & &\\
$B_c^- \to \psi^{'}K_1^-(1400)$ & $2.7 \times 10^{-5}$ & & & &\\
\hline\hline
  \end{tabular}
\end{center}
\end{table}

The most important sources of uncertainties for the branching ratios of the $B_c \to \eta_c^{'}M$ decays  come from the  $\beta_{B_c}$,  $\beta_{\eta_c^{'}}$, and   $\beta_{M}$ (with $M = D_{(s)}^{(*)}$) parameters, which are a relativistic correction  to the  wave function in the ISGW2 model,  the QCD coefficient $a_2$ (when $M = D_{(s)}^{(*)}$) and the decay constants $f_{\eta_c{'}}$, $f_{D_s}$, $f_{K_1(1270)}$ and $f_{K_1(1400)}$.   The dominant source of error come from the $\beta_{B_c}$ and  $\beta_{\eta_c^{'}}$ parameters, and the decay constant $f_{\eta_c^{'}} = (270 \pm 62)$ MeV \cite{CFW}. Moreover, the $B_c \to \eta_{c}^{'}M$ decays, with $M = D_s, \; K_1(1270), \; K_1(1400)$, are very sensitive to the decay constants  $f_{D_s} = 259$ \cite{CLEO}  $(241$ \cite{FDLS2008})  MeV, $|f_{K_1(1270)}|= 169.5 \;  ^{+18.8}_{-21.2}$ MeV, $|f_{K_1(1400)}|= 139.2 \; ^{+41.3}_{-45.6}$ MeV \cite{CC2010}, respectively.  On the other hand,  a variation of the $\beta_{B_c}$ and $\beta_{\eta_c{'}}$ parameters generates a greater increase in    $Br(B_c \to \eta_{c}^{'} D_{(s)}^{(*)})$ than in  $Br(B_c \to \eta_{c}^{'}M)$ when $M$ is  a charmless meson. In order to illustrate, we display in Table IV the variation of $Br(B_c^- \to \eta_c{'}D_s^-)$ in function of these input parameters. \\

For the $B_c \to \psi^{'}M$ modes, the most important sources of uncertainties come from the $\beta_{B_c}$, $\beta_{\psi^{'}}$, and   $\beta_{M}$ (with $M = D_{(s)}^{(*)}$) parameters, the relativistic correction to the form factor $f^{'}$ (which arises from corrections to the matrix elements of the axial vector current in the ISGW2 model) and the decay constants $f_{D_s}$, $f_{K_1(1270)}$ and $f_{K_1(1400)}$.  The dominant source of error comes from the relativistic correction to $f^{'}$ and the decay constants $f_{D_s}$, $f_{K_1(1270)}$ and $f_{K_1(1400)}$. For illustrating, we show in Table V the variation of $Br(B_c^- \to \psi{'}D_s^-)$ in function of these input parameters.  The $\beta_{B_c}$  parameter gives a bigger variation in $Br(B_c \to \eta_c^{'}M)$  than in $Br(B_c \to \psi^{'}M)$. In the same way,  the QCD parameter $a_2$ and the $\beta_{D_{(s)}^{(*)}}$ generate smaller variations in  $Br(B_c \to \psi^{'}D_{(s)}^{(*)})$ than in $Br(B_c \to \eta_c^{'}D_{(s)}^{(*)})$.\\

\begin{table}[ht]
{\small Table IV. $Br(B_c \to \eta_c^{'}D_s)$ in units  of  $10^{-5}$ in function of some input parameters.}
\par
\begin{center}
\renewcommand{\arraystretch}{1.5}
{\footnotesize
\begin{tabular}{c|c|c|c|c|c|c|c}
  \hline
    $\beta_{B_c}$ (GeV) & 0.86 & 0.88& 0.90& 0.92 & 0.94& 0.96&0.98 \\
 $Br(B_c \to \eta_c^{'}D_s)$  & 1.2 & 2.58 & 4.43 & 6.72 & 9.41 & 12.4 & 15.8 \\
 \hline\hline
 $\beta_{\eta_c^{'}}$  (GeV) & 0.82 & 0.84 & 0.86 & 0.88 & 0.90 & 0.92& 0.94 \\
 $Br(B_c \to \eta_c^{'}D_s)$ & 18.3 & 13.8 & 10& 6.72 & 4.08 & 2.09 & 0.76\\
 \hline\hline
 $\beta_{D_s}$ (GeV) & 0.50 & 0.52 & 0.54 & 0.56 & 0.58 & 0.60 & 0.62 \\
 $Br(B_c \to \eta_c^{'}D_s)$ & 8.19 & 7.67 & 7.18 & 6.72 & 6.29 & 5.88 & 5.50\\
 \hline\hline
 $a_2$ & $-0.17$ & $-0.18$ & $-0.19$ & $-0.20$ & $-0.21$ & $-0.22$ & $-0.23$ \\
 $Br(B_c \to \eta_c^{'}D_s)$ & 8.65 & 7.98 & 7.34 & 6.72 & 6.13 & 5.57 & 5.04 \\
 \hline\hline
 $f_{\eta_c^{'}}$ (GeV) & 0.208 & 0.270 & 0.332 & & & & \\
 $Br(B_c \to \eta_c^{'}D_s)$ & 9.77 & 6.72 & 4.24  & & & & \\
 \hline\hline
 $f_{D_s}$ (GeV) & 0.241 & 0.259 & & & & & \\
  $Br(B_c \to \eta_c^{'}D_s)$ & 5.07 & 6.72 & & & & & \\
  \hline\hline
 \end{tabular}
}
\end{center}
\end{table}

\begin{table}[ht]
{\small Table V. $Br(B_c \to \psi^{'}D_s)$ in units  of  $10^{-4}$ in function of some input parameters.}
\par
\begin{center}
\renewcommand{\arraystretch}{1.5}
{\footnotesize
\begin{tabular}{c|c|c|c|c|c|c|c}
  \hline
    $\beta_{B_c}$ (GeV) & 0.86 & 0.88& 0.90& 0.92 & 0.94& 0.96 & 0.98 \\
 $Br(B_c \to \psi^{'}D_s)$  & 4.9 & 5.03 & 5.14  & 5.25  & 5.34  & 5.41  & 5.48  \\
 \hline\hline
 $\beta_{\psi^{'}}$  (GeV) & 0.56 & 0.58 & 0.60 & 0.62 & 0.64 & 0.66 & 0.68  \\
 $Br(B_c \to \psi ^{'}D_s)$ & 5.51 & 5.46 & 5.37  & 5.25 & 5.1 & 4.9 & 4.7 \\
 \hline\hline
 $\beta_{D_s}$ (GeV) & 0.50 & 0.52 & 0.54 & 0.56 & 0.58 & 0.60 & 0.62 \\
 $Br(B_c \to \psi^{'}D_s)$ & 5.21 & 5.22 & 5.24  & 5.25 & 5.26 & 5.27 & 5.28 \\
 \hline\hline
 $f_{rel}^{'}$ &  0.810 &  0.855 & 0.881 & 0.899 & 0.907 & 0.943 & 0.988 \\
 $Br(B_c \to \psi^{'}D_s)$  & 4.24 & 4.73 & 5.03 & 5.25 & 5.34 & 5.79 & 6.37 \\
 \hline\hline
 $a_2$ & $-0.17$ & $-0.18$ & $-0.19$ & $-0.20$ & $-0.21$ & $-0.22$ & $-0.23$ \\
 $Br(B_c \to \psi^{'}D_s)$ & 5.21 & 5.22  & 5.23  & 5.25  & 5.26 & 5.27 & 5.29  \\
 \hline\hline
  $f_{D_s}$ (GeV) & 0.241 & 0.259 & & & & & \\
  $Br(B_c \to \psi^{'}D_s)$ & 4.56 & 5.25  & & & & & \\
  \hline
 \end{tabular}
}
\end{center}
\end{table}

For completeness, we have computed  the branching ratios for the semileptonic $B_c \to \eta_c^{'}(\psi^{'})e\nu_e$\footnote{Decay widths of the $B_c \to \eta_c^{'}(\psi^{'})e\nu_e$ processes were calculated in Ref. \cite{isgw2}. So, in this case we  obtained simply these numerical values using updated inputs.} and $B_c \to \eta_c^{'}(\psi^{'})\tau \nu_{\tau}$ decays. In Table VI,  we show our results and compare with predictions in other approaches based on  QCD sum rules \cite{WL2008,kiselev2003}, relativistic \cite{EFG2003,CF2000,LT1997} and nonrelativistic \cite{ChCh1994} quark models. In general,  predictions for $Br(B_c^- \to \eta_c^{'}e^-\overline{\nu}_e)$  are of the order of $10^{-4}$ in the different approaches except in the framework of the light-cone QCD sum rules approach \cite{WL2008}, where it is obtained the biggest value.  For the $B_c^- \to \eta_c^{'}\tau^-\overline{\nu}_{\tau}$ decay, our result is the smallest but  close to numerical value of Ref. \cite{kiselev2003}.  The prediction obtained in Ref.
\cite{WL2008} is the biggest. It is six times our numerical value.\\

On the other hand,  our prediction for the   branching ratio of the $B_c^- \to \psi^{'}e^-\overline{\nu}_e$ decay is the biggest. It is of the order of $10^{-3}$. A similar result is obtained in Refs. \cite{CF2000,ChCh1994,kiselev2003}. For  $B_c^- \to \psi^{'}\tau^-\overline{\nu}_{\tau}$ channel, we compute the branching ratio using the expression for $d\Gamma(B_c \to V\tau \nu)/dq^2$ displayed in  Ref. \cite{WSL2009}. Our prediction for the branching of this process is $\sim$ two times the numerical value  of Ref. \cite{kiselev2003}. We obtained for the three kinds of the $B_c \to \psi^{'} \tau \overline{\nu}_{\tau}$ decays that  the longitudinal $(\Gamma_L)$ and transverse $(\Gamma_T)$ contributions are
\begin{eqnarray}\nonumber
\Gamma_L &=&  6.6 \times 10^{-5}, \nonumber \\
\Gamma_T &=& 8.2 \times 10^{-5},
\end{eqnarray}

 \emph{i.e.}, $\Gamma_L$ is comparable with $\Gamma_T$.  For this process, the ratio $\Gamma_L/\Gamma_T$ is $0.8$. A similar result was presented in Ref.  \cite{WSL2009} for the $B_c \to J/\psi \tau \overline{\nu}_{\tau}$ mode.\\

Finally, from our numerical values showed in Table VI, we get the following ratios:
\begin{equation}\nonumber
\frac{Br(B_c^- \to \eta_c^{'}e^-\overline{\nu}_e)}{Br(B_c^- \to \eta_c^{'}\tau^-\overline{\nu}_{\tau})} = 35.4 \ \ \ \ \text{and}  \ \ \ \  \frac{Br(B_c^- \to \psi^{'}e^-\overline{\nu}_e)}{Br(B_c^- \to \psi^{'}\tau^-\overline{\nu}_{\tau})} = 14.
\end{equation}
The first quotient is too big. In Refs. \cite{WL2008} and \cite{kiselev2003} it is obtained $13.5$ and $12.5$, respectively, while predictions of Ref. \cite{kiselev2003} give 11.75 for the second ratio. We also compute the following quotients:
\begin{equation}\nonumber
\frac{Br(B_c^- \to \psi^{'}e^-\overline{\nu}_e)}{Br(B_c^- \to \eta_c^{'}e^-\overline{\nu}_e)}=   4.6          \ \ \ \   \text{and}       \ \ \ \   \frac{Br(B_c^- \to \psi^{'}\tau^-\overline{\nu}_{\tau})}{Br(B_c^- \to \eta_c^{'}\tau^-\overline{\nu}_{\tau})} = 11.5.
\end{equation}
Results of Ref. \cite{kiselev2003} give 4.7 and 5, respectively, for these ratios. Our prediction for the second quotient is $\approx$ two times the numerical value obtained from the Ref. \cite{kiselev2003}.
\\

\begin{table}[ht]
{\small Table VI. Branching ratios of the semileptonic $B_c \to \eta_c^{'}(\psi^{'})l\nu$ decays.}
\par
\begin{center}
\renewcommand{\arraystretch}{1.5}
{\footnotesize
\begin{tabular}{c|c|c|c|c|c|c|c}
  \hline
    Decay & This work & \cite{WL2008}& \cite{kiselev2003}&\cite{EFG2003} & \cite{CF2000}& \cite{LT1997}&\cite{ChCh1994} \\
    \hline\hline
 $B_c^- \to \eta_c^{'}e^-\overline{\nu}_e$ & $4.6 \times 10^{-4}$ & $1.1 \times 10^{-3}$ &$2 \times 10^{-4}$ & $3.2 \times 10^{-4}$ &$2.1 \times 10^{-4}$ &$4.2 \times 10^{-4}$ &$5 \times 10^{-4}$ \\
 $B_c^- \to \eta_c^{'}\tau^-\overline{\nu}_{\tau}$ &$1.3 \times  10^{-5} $ & $8.1 \times 10^{-5}$ & $1.6 \times 10^{-5}$ & & & & \\
 \hline\hline
 $B_c^- \to \psi^{'}e^-\overline{\nu}_e$ & $ 2.1 \times 10^{-3}$& & $9.4 \times 10^{-4}$ & $3 \times 10^{-4}$ & $1.2 \times 10^{-3}$ &$1.3 \times 10^{-4}$ & $1 \times 10^{-3}$ \\
 $B_c^- \to \psi^{'}\tau^-\overline{\nu}_{\tau}$ &$1.5 \times 10^{-4}$ & & $8 \times 10^{-5}$ & & & &\\
 \hline\hline
  \end{tabular}
}
\end{center}
\end{table}

\bigskip

\section{Conclusions}

In this work we   studied in a systematic way the production of radially excited charmonium mesons in two-body nonleptonic $B_c$ decays assuming factorization approach and using the  ISGW2 quark model \cite{isgw2}, which is an improved version of the nonrelativistic ISGW model \cite{isgw}. We  obtained  branching ratios for $B_c \to X_{c\overline{c}}(2S)M$ decays, where $X_{c\overline{c}}(2S)$ is the $\eta_c'$ or  $\psi'$ meson, and $M$ is a pseudoscalar ($P$) or a vector ($V$) or an axial-vector ($A(^3P_1)$) meson. We  compared our predictions with previous results obtained in  other approaches and gave some ratios that could be an additional  test for the different frameworks used for calculating these branching ratios. We found that some of these decays have branching ratios of the  order of $10^{-3} - 10^{-4}$, which indicates that they could be measured in the future at LHCb experiment. For completeness we computed branching ratios of semileptonic $B_c \to \eta_c{'}(\psi^{'})l\nu$ decays and compared with results obtained in other scenarios.\\

$Br(B_c \to \psi^{'}M)$ is very sensitive to the relativistic correction to the form factor  $f{'}$ while $Br(B_c \to \eta_c^{'}M)$  to the   $\beta_{B_c}$ and $\beta_{\eta_c^{'}}$ parameters (which  are also relativistic corrections).  Although the ISGW2 model includes   relativistic corrections to the matrix elements of the axial vector current and the  wave functions through the effective interquark potential,  the branching ratios obtained in this model are much larger than theoretical predictions in relativistic quark models \cite{EFG2003,CF2000,LT1997}. This could indicate that the relativistic effects on these $B_c$ decays are not negligible. Therefore, the comparison of the two-body nonleptonic $B_c$ decays with radially excited charmonium mesons in the final states among different theoretical model predictions may also help in understanding the relativistic effects on the exclusive $B_c$ decays.\\

Our main results are:
\begin{itemize}
\item For $B_c \to \eta_c'M$ decays, the branching ratios of the CKM favored $B_c^- \to \eta_c' \pi^-(\rho^-), \; \eta_c'a_1^-$,  $\eta_c'D_s^*$ modes are of the order of $10^{-4}$. We find that the behavior of the interference term in $B_c^- \to \eta_c'D(D_s)$ and $B_c \to \eta_c'D^*(D_s^*)$ decays is different. In the first case, it is negative while in the second case it is positive because the form factor $A_2(t=m^2_{\eta_c'})$ and the QCD coefficient $a_2$ are negative.
\item For $B_c \to \psi^{'}M$ decays, our predictions are the biggest. The branching ratios of the CKM favored $B_c \to \psi^{'}\rho^-, \; \psi^{'}a_1^-, \; \psi^{'}D_s^{*-}$ channels are of the order of $10^{-3}$. The branching ratio of the exclusive $B_c \to \psi^{'}a_1^-$ decay is the biggest.
 \item For the semileptonic $B_c^- \to \psi^{'}\tau\overline{\nu}_{\tau}$ we obtain that the longitudinal ($\Gamma_L$) and transverse ($\Gamma_T$) contributions are $8.2 \times 10^{-5}$ and $6.6 \times 10^{-5}$, respectively. So, the ratio $\Gamma_T/\Gamma_L$ is 0.8.
\end{itemize}

\bigskip

\begin{center}
\textbf{Acknowledgements}
\end{center}

This work has been partly supported by CNPq (Brazil) and Comit\'e Central de Investigaciones of Universidad del Tolima.

\end{document}